\documentclass{emulateapj} 
\usepackage{apjfonts}
\usepackage{psfig}
\usepackage{mathrsfs}
\begin{document}
%\received{}
%\accepted{}
%\revised{}
\slugcomment{AJ, in press}

\title{Image-Subtraction Photometry of Variable Stars\\ in the Globular 
Clusters NGC~6388 and NGC~6441} 
\author{T.~Michael~Corwin, Andrew N. Sumerel} 
\affil{Department of Physics and Optical Science, University of North Carolina 
at Charlotte, Charlotte, NC 28223}
\email{mcorwin@uncc.edu, ansumere@uncc.edu}   

\author{Barton J. Pritzl\altaffilmark{1}}
\affil{Department of Physics and Astronomy, Macalester College, 1600 Grand Avenue, Saint Paul, MN 55105} 
\email{pritzl@macalester.edu}

\author{Horace A. Smith}
\affil{Department of Physics and Astronomy, Michigan State University,
       East Lansing, MI 48824}
\email{smith@pa.msu.edu}
 \author{M.~Catelan}
\affil{Pontificia Universidad Cat\'olica de Chile, Departamento de
       Astronom\'\i a y Astrof\'\i sica, \\ Av. Vicu\~{n}a Mackenna 4860,
      782-0436 Macul, Santiago, Chile}
\email{mcatelan@astro.puc.cl}

\author{Allen V. Sweigart}
\affil{NASA Goddard Space Flight Center, Exploration of the Universe
Division, Code 667, Greenbelt, MD 20771.}
\email{sweigart@bach.gsfc.nasa.gov}

\and

\author{Peter B. Stetson}
\affil{Dominion Astrophysical Observatory, Herzberg Institute of Astrophysics,
National Rsearch Council of Canada,\\ 5071 West Saanich Road, Victoria, BC,
V9E 2E7, Canada}
\email{Peter.Stetson@nrc-cnrc.gc.ca}

\altaffiltext{1}{Visiting Astronomer, Cerro Tololo Inter-American
Observatory, National Optical Astronomy Observatories, which is
operated by AURA, Inc., under cooperative agreement with the
National Science Foundation}

\begin{abstract}
We have applied Alard's image subtraction method (ISIS~v2.1)  
to the observations of the globular clusters NGC~6388 and
NGC~6441 previously analyzed using standard photometric techniques
({\sc daophot, allframe}). 
In this reanalysis of observations obtained at CTIO, besides recovering
the variables previously detected on the basis of our ground-based images, 
we have also been able to recover most of 
the RR Lyrae variables previously detected only in the analysis of 
{\em Hubble Space Telescope} WFPC2 observations of the inner region of 
NGC~6441. In addition, we report five possible new variables not found in 
the analysis of the HST observations of NGC~6441. This dramatically 
illustrates the capabilities of image subtraction techniques applied to 
ground-based data to recover variables in extremely crowded fields. 
We have also detected twelve new variables and six possible variables 
in NGC~6388 not found in our previous ground-based studies. 
The revised mean period for RRab stars in NGC~6388 is 0.676 day, while
the mean period of RRab stars in NGC~6441 is unchanged at 
0.759 day.  These values are among the
largest known for any galactic globular cluster. Additional probable type II
Cepheids were identified in NGC~6388, confirming its status as a 
metal-rich globular cluster rich in Cepheids.

\end{abstract}


  \section{Introduction} 
\citet{R97}
discovered that the horizontal branch morphologies of
NGC~6388 and NGC~6441 were different from those of other metal-rich globular
clusters.  Although these two clusters have metallicities near 
${\rm [Fe/H]} = -0.6$
\citep{A88, Cl05}, they have blue extensions to the horizontal branch in
addition to the red horizontal branch components usually seen in metal-rich globular clusters.
As a consequence, and unlike other metal-rich globular clusters, NGC~6388 and
NGC~6441 have substantial populations of RR Lyrae stars.  These RR Lyrae stars
are themselves distinguished by having extraordinarily long periods for their
metallicities, so that they do not fit the 
usual pattern of decreasing mean period versus
increasing [Fe/H] \citep{L99,P00,P01,P02,P03}.
The reasons for the unusual horizontal branch morphologies of these clusters and the
unusual characteristics of their RR Lyrae stars have not
been established, although several scenarios have been advanced \citep{P97,S98,S02,
R02, Ca05}.  In addition to harboring
this anomalous population of RR Lyrae stars, NGC~6388 and NGC~6441 each contain
several type II Cepheids, making them the most metal-rich globular clusters known to
contain such stars \citep{P02,P03}.

%                                                One column figure
%----------------------------------------------------------- Fig. 1
\begin{figure*}[t]
  \figurenum{1}
  \epsscale{0.85}
 \plotone{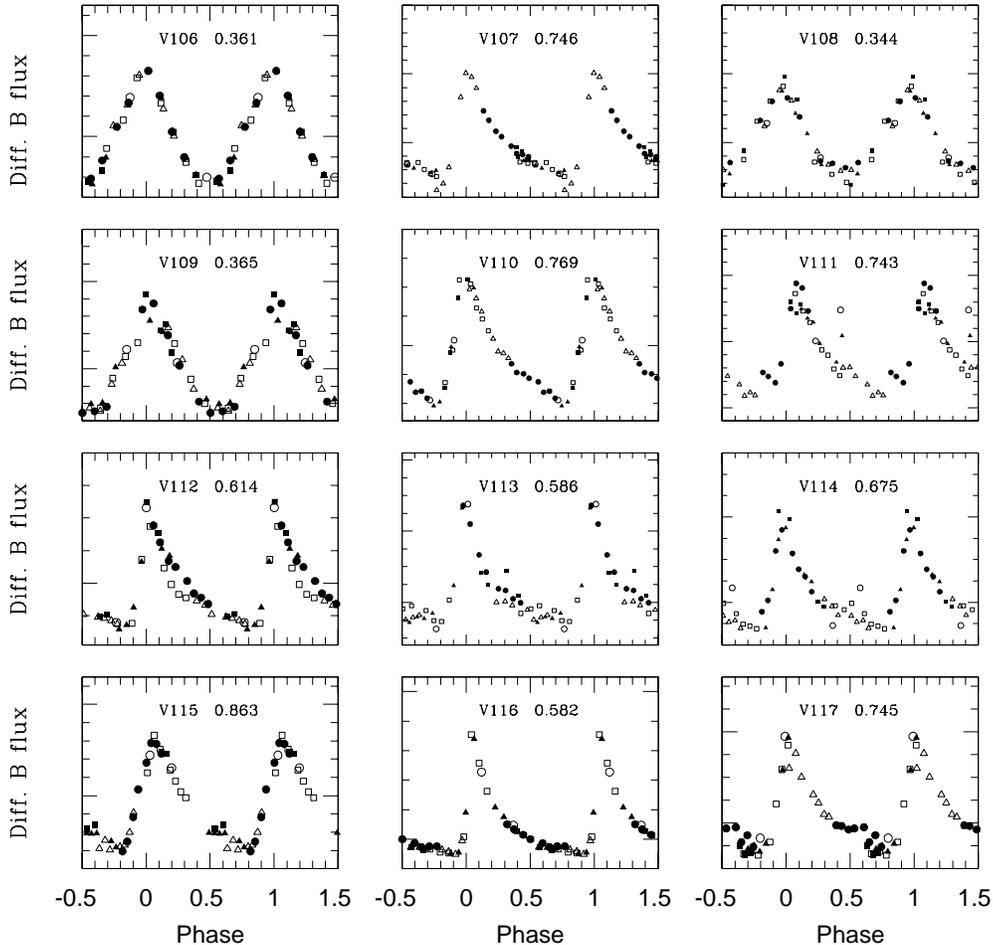}
  \caption{Differential $B$ flux light curves for the 24 NGC~6441 variable 
stars not found in \citet{P01}, but found in the HST data of
\citet{P03}. The data range from HJD 2450959 to 2450968. The order of the data is 
filled squares (nights 1 and 3), open squares (night4), filled triangles (night 
7), open triangles (night 8), filled circles (night 9), and open circles (night 
10). Differential fluxes are in arbitrary linear units.
      }
      \label{Fig01a}
\end{figure*}

%                                                One column figure
%----------------------------------------------------------- Fig. 1 (continued) 
\begin{figure*}[t]
  \figurenum{1}
  \epsscale{0.85}
 \plotone{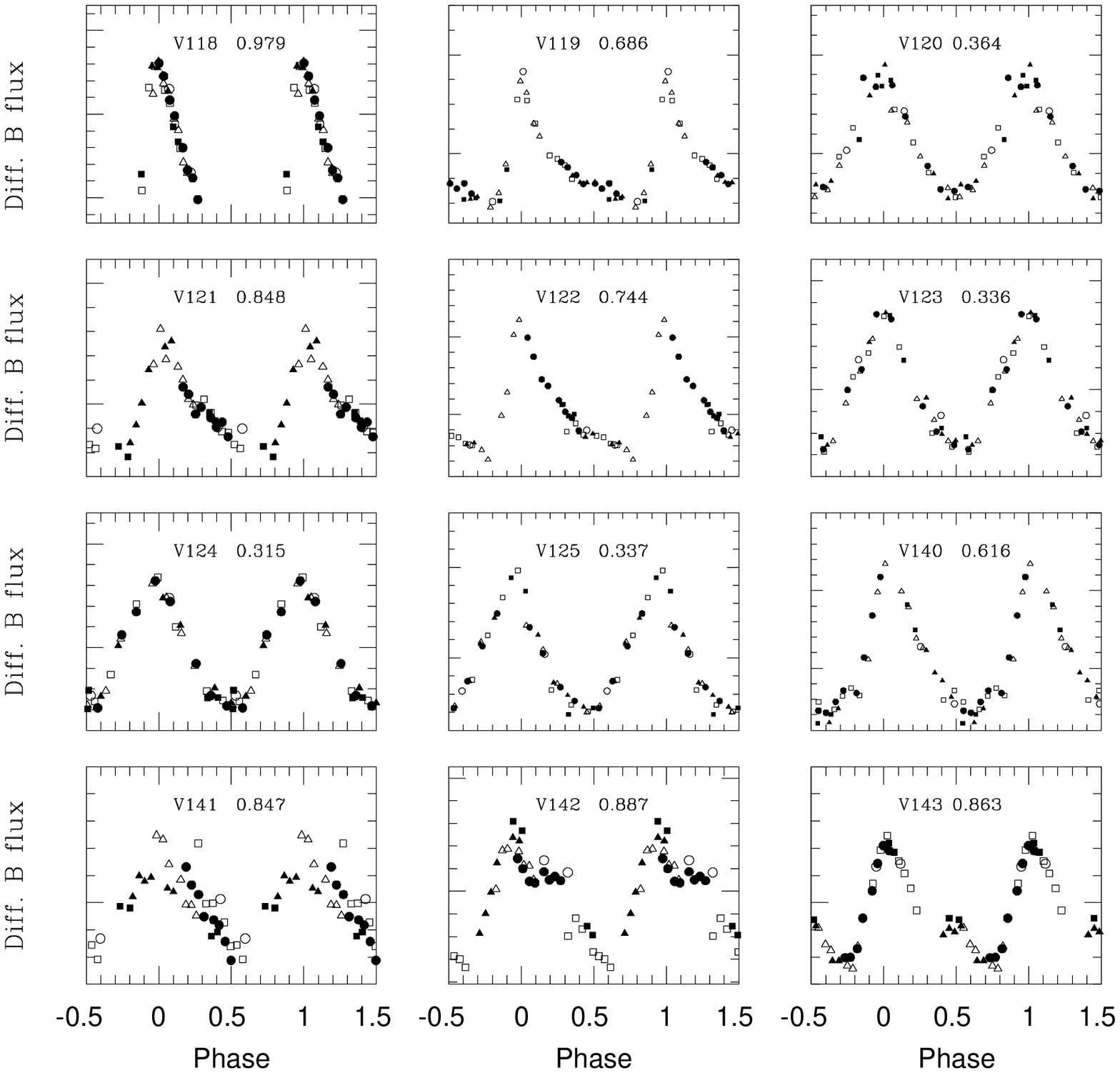}
  \caption{{\em Continued.}
      }
      \label{Fig01b}
\end{figure*}

%                                                One column figure
%----------------------------------------------------------- Fig. 2  
\begin{figure*}[t]
  \figurenum{2}
  \epsscale{0.85}
\plotone{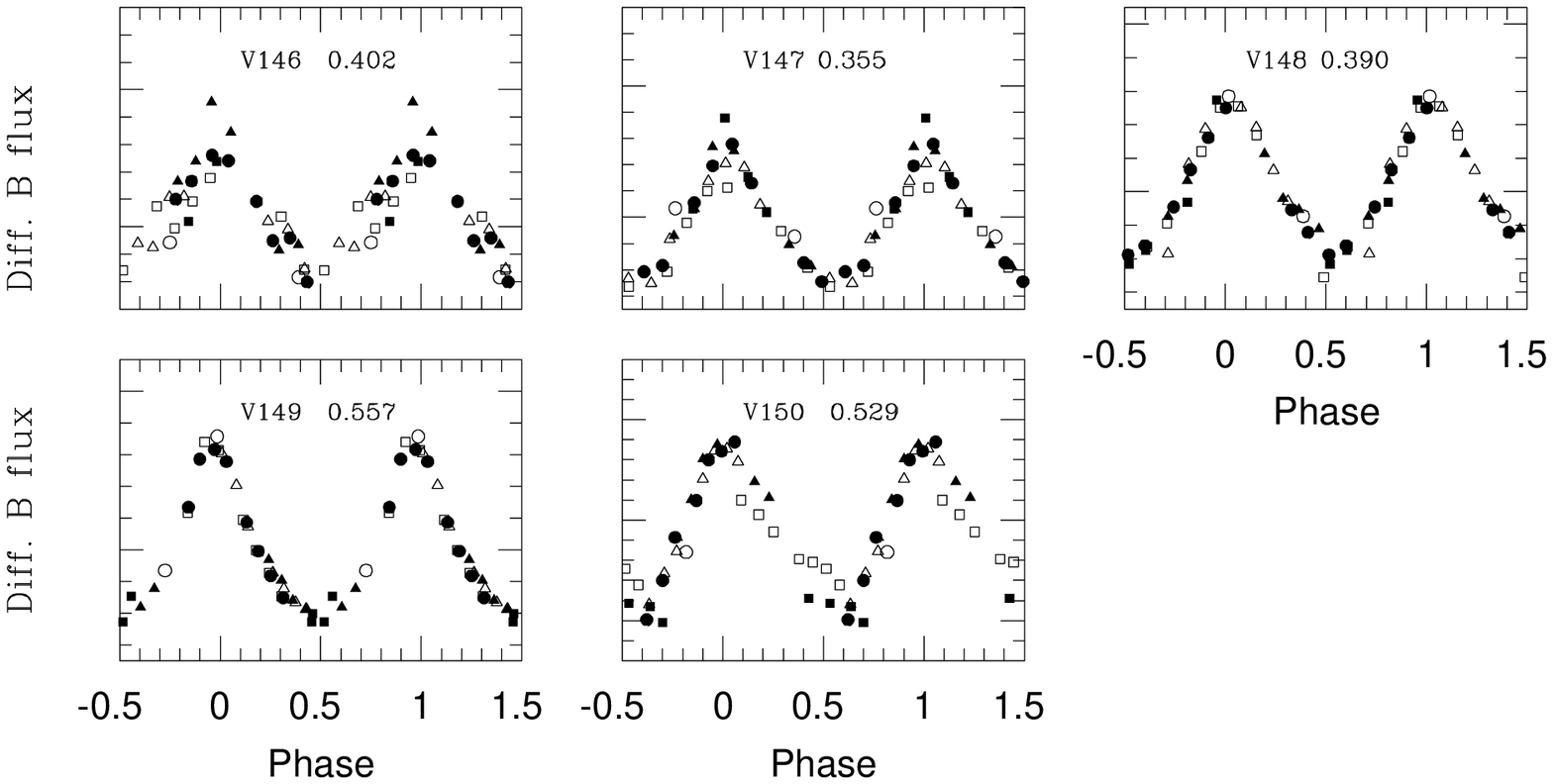}
  \caption{Differential $B$ flux light curves for the five NGC~6441 variable 
stars found in neither \citet{P01} nor \citet{P03}. The data range and 
symbols are as in Figure~1.
      }
      \label{fig2}
\end{figure*}

\citet{P01,P02} used {\sc daophot} \citep{S87} and {\sc allframe} \citep{S94} to analyze CCD images
of NGC~6388 and NGC~6441 that were obtained with the 0.9-m telescope at CTIO.  Many
variable stars were identified in these studies, but the completeness of the variable
star searches was low in the most crowded central portions of the clusters. 
On the other hand, 
image-subtraction techniques have revealed large numbers of variable stars that had
previously gone undetected on the basis of more standard techniques,
including {\sc allframe} (\citeauthor{C03} \citeyear{C03}, \citeyear{Co04} 
and references therein).  
We have therefore reanalyzed the 0.9-m observations using the ISIS v2.1
image-subtraction package \citep{A00,A98} in
order to obtain a more complete inventory of the variable star 
populations of the clusters.
The utility and completeness of the image-subtraction method can
itself be evaluated in the case of NGC~6441, since \citet{P03} also studied the
variable stars of the inner regions of that cluster using snapshot observations obtained
with the WFPC2 camera on the {\em Hubble Space Telescope} (HST).

%                                                One column figure
%----------------------------------------------------------- Fig. 3  
\begin{figure*}[t]
  \figurenum{3}
  \epsscale{0.85}
\plotone{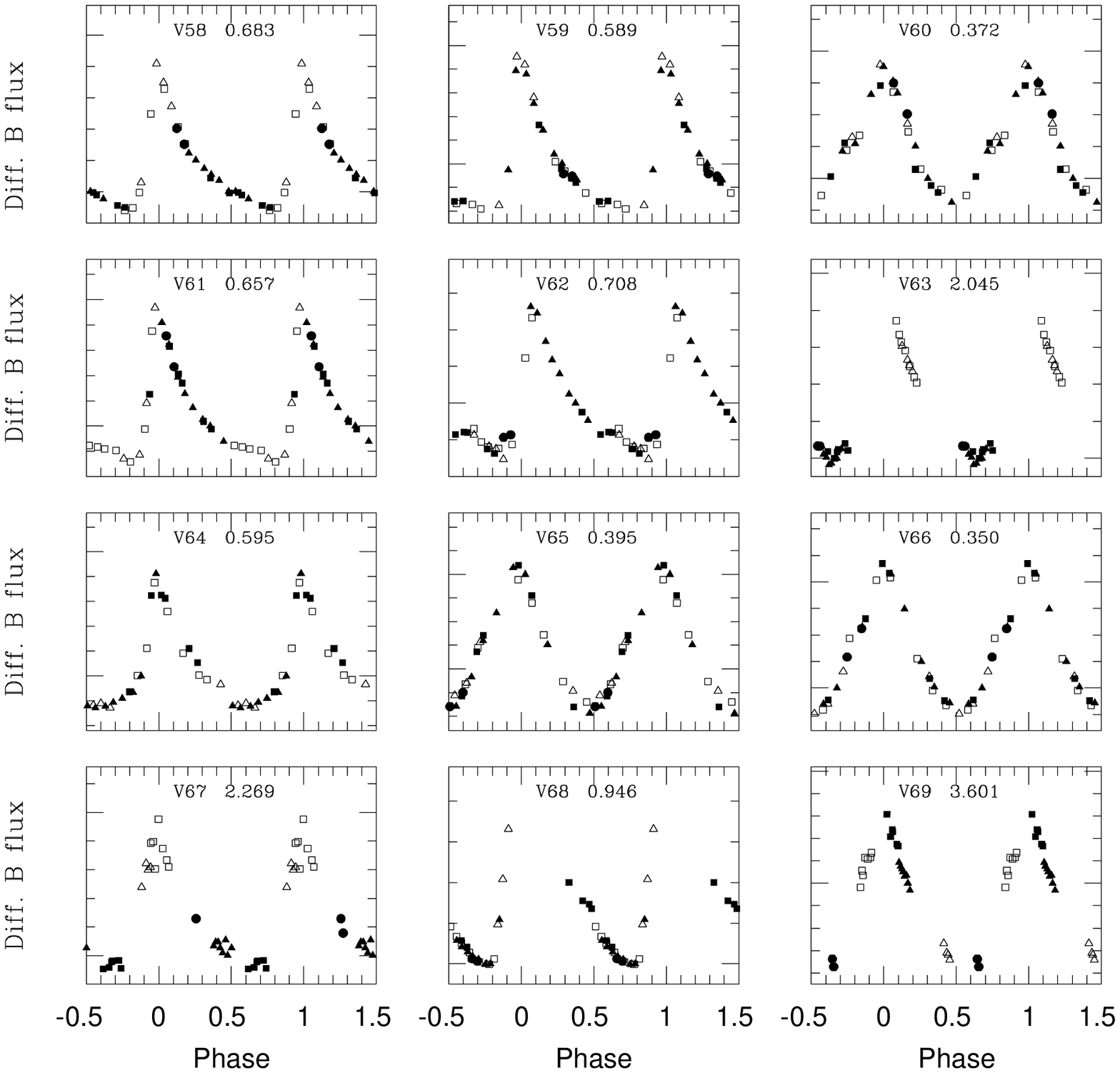}
  \caption{Differential $B$ flux light curves for the twelve NGC~6388 variable 
stars not found in \citet{P02}. The data range and symbols are as in 
Figure~1. 
      }
      \label{fig3}
\end{figure*}

%                                                One column figure
%----------------------------------------------------------- Fig. 4 
\begin{figure*}[t]
  \figurenum{4}
  \epsscale{0.85}
\plotone{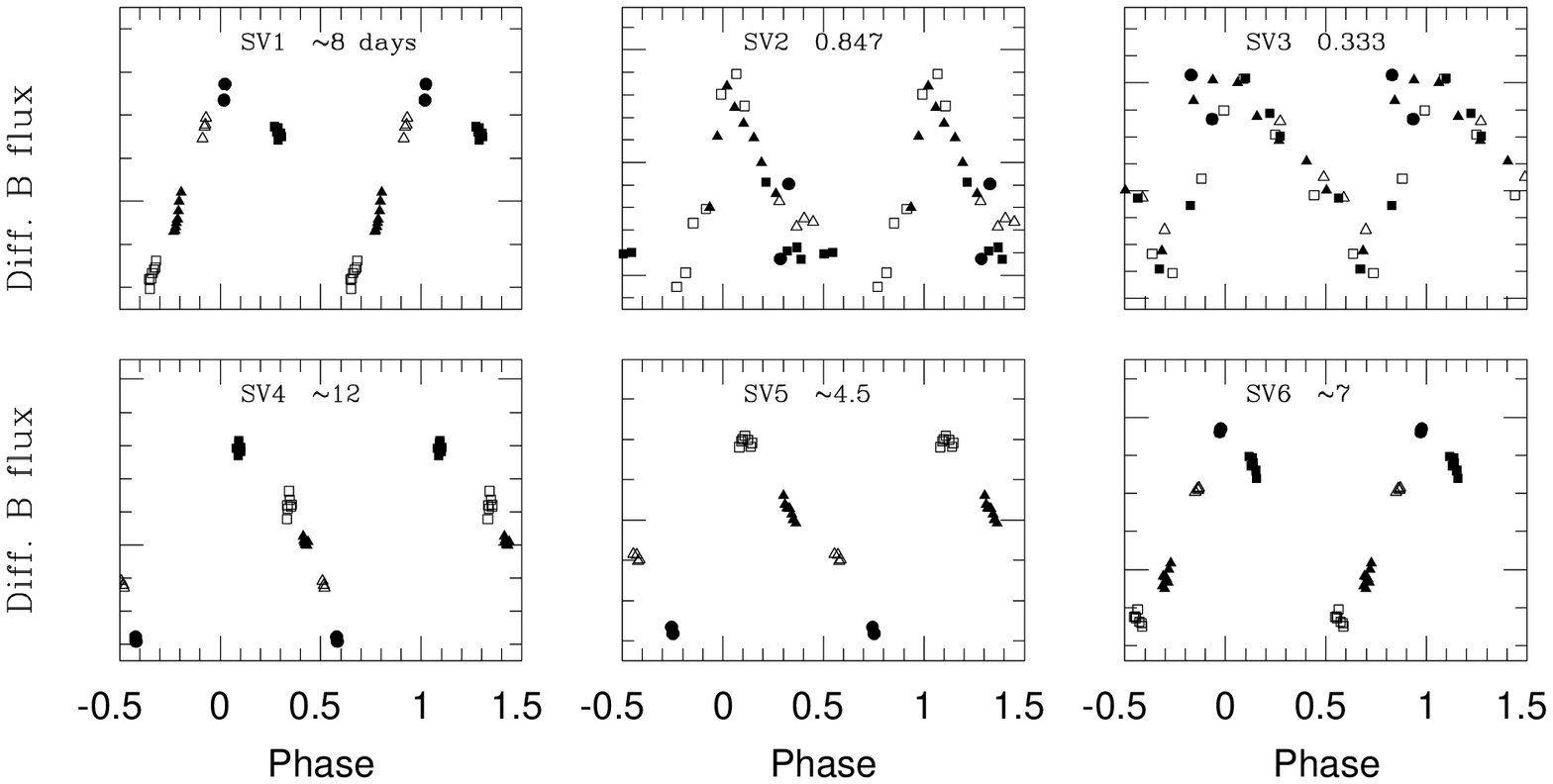}
  \caption{Differential $B$ flux light curves for the six
suspected NGC~6388 variables. The data range and symbols are as in 
Figure~1. 
      }
      \label{fig4}
\end{figure*}

\section{Observations and Reductions}
 The details of the observations and processing of the images can be found in 
\citet{P01, P02}. The ISIS analysis measures the 
difference in flux for pixels in each image of 
the time series relative to their flux in a reference image obtained by 
stacking a suitable subset of images.  In the ISIS method, the original
images are convolved
with a kernel to account for seeing
variations and geometrical distortions 
of the individual frames. We used the 10 $B$ images with the best seeing 
from
each of the NGC~6441 and the NGC~6388 datasets to build up the reference images.
This process identified a large number of stars with differential flux
values above our threshold.  Each was evaluated for the likelihood that it
was a variable star and most clearly were not.  We then compared our list
of possible variables with the list of known variables.  Any of our
suspect variables that were not previously known variables were closely
examined and many were eliminated.  Some appeared to be genuinely variable
and are reported here as new variables.

\begin{deluxetable}{llll}                                        
\tablecaption{NGC~6441 HST Variables\label{tbl-1}}
\tablewidth{0pt}
\tablehead{
\colhead{Variable}  & \colhead{ISIS period}  & \colhead{HST period} & 
\colhead{Type}}
\startdata
V106    &0.361  &0.36092 &RRc\\ 
V107    &0.746  &0.73891 &RRab\\  
V108    &0.344  &0.34419 &RRc\\  
V109    &0.365  &0.36455 &RRc\\ 
V110    &0.769  &0.76867 &RRab\\ 
V111    &0.743  &0.74464 &RRab\\ 
V112    &0.614  &0.61419 &RRab\\ 
V113    &0.586  &0.58845 &RRab\\ 
V114    &0.675  &0.67389 &RRab\\ 
V115    &0.860  &0.86311 &RRab\\ 
V116    &0.582  &0.58229 &RRab\\ 
V117    &0.745  &0.74529 &RRab\\ 
V118    &0.979  &0.97923 &RRab or Ceph\\ 
V119    &0.686  &0.68628 &RRab\\ 
V120    &0.364  &0.36396 &RRc\\ 
V121    &0.848  &0.83748 &RRab\\ 
V122	&0.744	 &0.74270 &RRab\\
V123	&0.336	 &0.33566 &RRc\\
V124	&0.315  &0.31588 &RRc\\	
V125	&0.337	 &0.33679 &RRc\\
V140	&0.616  &0.35180 &RR\\
V141	&0.847	 &0.84465 &RRab\\
V142	&0.887	 &0.88400 &RRab\\
V143	&0.863	 &0.86277 &RRab\\
\enddata 
\end{deluxetable}

\begin{deluxetable}{lllll}
\tablecaption{NGC~6441 Possible New Variables\label{tbl-2}}
\tablewidth{0pt}
\tablehead{
\colhead{Variable}  & \colhead{ISIS period} & \colhead{RA (2000)} & 
\colhead{Dec (2000)} & \colhead{Type}}
\startdata
NV1(V146)	&0.402	 &17 50 13.15&-37 03 00.4&RRc\\
NV2(V147)	&0.355	 & 17 50 13.26&-37 02 52.3&RRc\\
NV3(V148)	&0.390	 &17 50 12.79&-37 02 50.9 &RRc\\
NV4(V149)	&0.557	 & 17 50 10.06& -37 02 26.5&RR\\
NV5(V150)	&0.529	 &17 50 07.07 &-37 03 16.1 &RR\\
\enddata
\end{deluxetable}

\section{Results}

In this section we present results for newly discovered variable stars in both
NGC~6388 and NGC~6441.  In the case of NGC~6441, we also evaluate the ability
of ISIS to detect RR Lyrae and Cepheid variables in ground-based data.  
In particular,
we compare the detection of variable stars in the ISIS analysis of images of
NGC~6441 taken at CTIO to the detection of variable stars in  
\citeauthor{P03}~'s (\citeyear{P03})
analysis of WFPC2 observations of the same cluster.

Periods for the new variables were determined using the period-search 
program {\sc kiwi}. {\sc kiwi} follows \citet{lk65}
in searching for periodicity by seeking 
to minimize the total length of the line segments that join adjacent observations 
in phase space, i.e., to maximize the smoothness of the light curve. 
(The {\sc kiwi} 
program was kindly provided to us by Dr. Betty Blanco.) In some cases 
the {\sc kiwi} periods were 
adjusted to improve the phase match between different nights.  
The analyzed data cover 10 nights, spanning about 33 cycles for the shorter-period 
variables, and about 12 cycles for the longer-period ones. Because of this 
relatively short time interval, ISIS periods are given to only three significant 
figures.   Differential flux light curves for the variables based 
on the ISIS
 analysis and the periods given in Tables~1 and
 2 (NGC~6441) and 4 (NGC~6388) are shown in Figures~1 and 2 (NGC~6441)
  and 3 and 4 (NGC~6388). 
  
Differential fluxes can be transformed into standard magnitudes if
reliable photometry can be obtained for the variable stars in the
reference frame \citep[see, for example,][]{Br03,Mo02,Ba05}.  
Unfortunately, the newly discovered variables in NGC~6388 and NGC~6441
are in very crowded portions of the field, where the star images are
badly blended.  Conventional methods of photometry such as {\sc allstar}
and {\sc allframe} are in these cases unable to provide the accurate
magnitude zero-point on which to base a conversion from differential
fluxes to magnitudes.  Thus, our analysis will be based upon
the differential flux light curves produced by ISIS.

Table~1 and Figure~1 refer to 
NGC~6441 variables previously identified with HST, whereas Table~2 and 
Figure~2 report on the new NGC~6441 variable candidates detected with ISIS. 
There is only one instance 
where the {\sc kiwi} period differs significantly from that found in \citet{P03}
(V140, see section 4). Table 2 contains five possible 
NGC~6441 variables that were not identified in the HST study by \citet{P03}. The WFPC2
photometry was reexamined to determine whether these five variables could be recovered
from the HST dataset.  All were recovered and could be identified as probable or
possible variables.  For NV1 and NV2 the WFPC2 photometry is noisy, but
is consistent with the periods identified from the ISIS data.  The WFPC2 photometry
for NV3 is especially noisy, but suggests that the star may be truly variable.
For NV4 and NV5, there are only 10 and 6 HST epochs of observation, respectively, but the
data are again consistent with the identification of NV4 and NV5 as variable
stars.  Thus, NV1 through NV5 would become NGC~6441 variables V146 through
V150, respectively.
The positions of these variables were determined from
the WFPC2 images as in \citet{P03}.

Three variables found in the HST analysis of \citet{P03} were not found in the ISIS 
analysis: V136, V138, and V145. All three of these variables are within 10 
arcsec of the center of the cluster. In fact, V145 is the variable found by HST 
that is closest to the center. It would be expected that stars near the very 
center of a cluster would be the most difficult for ISIS to detect.
It is
noteworthy in this connection that NGC 6388 and NGC 6441 are both clusters
with strong central concentrations of stars
\citep[see][]{H96, Tr95}.

An RRc classification for the newly discovered variables NV1, NV2, and
NV3 seems clear on the basis of period and light curve shape.  However,
results for the longer period variables NV4 and NV5 are not so clear.
Plots of the Fourier decomposition parameters of light curves
have proven useful in distinguishing between RRab and RRc type variables, 
e.g. \citet{P02}.  In Figure 5 we plot the Fourier parameters
$A_{21}$ versus $\phi_{21}$, based upon a fifth order fit
to the differential $B$ light curves. NV4 and NV5, as well as
V140, fall between the stars clearly established as belonging to 
Bailey type ab and the stars clearly established as Bailey type c.
We have also calculated the \citet{St87} skewness
parameter from the differential light curves.  This parameter is
defined as Sk = [1/(rise time in phase units)] - 1. Skewness
is plotted against $\phi_{21}$ in Figure 6.  Here, V140 and NV4
fall among the RRc variables, whereas the location of NV5 is closer
to those of the RRab stars.
The skewness and Fourier
parameters for NGC~6441 variables are listed in Table 3. A few
outlying points in the light curves were omitted in calculating the
parameters. Parameters were not calculated for V118 because of the large
gap in its observed light curve.

For NGC~6388, for which no HST variable star analysis has been
  completed, 
  Table~4 lists variable stars not previously detected 
  in the \citet{P02} study.  Light curves for the new variables NV1 through
  NV12 are shown in Figure~3,  whereas Figure~4 shows light 
  curves for additional suspected variables of uncertain variability 
  type. 

%\clearpage

\begin{deluxetable}{llllllll}                                    
\tablecaption{Fourier and Skewness Parameters for NGC~6441 Variables\label{tbl-4}}
\tablewidth{0pt}
\tablehead{
\colhead{Variable}  & \colhead{$A_{21}$}  & \colhead{$A_{31}$} & 
\colhead{$A_{41}$} & \colhead{$\phi_{21}$} & \colhead{$\phi_{31}$} & 
\colhead{$\phi_{41}$} & \colhead{Sk} }
\startdata
V106  & 0.13 &  0.05 & 0.08 & 3.21  &  5.34  &  3.07  & 1.49\\
V107 & 0.54  & 0.32  & 0.13  & 3.99  &  1.92  &  6.08 &  3.69\\
V108 & 0.15 &  0.10 & 0.05 & 2.99 &   4.93  &  2.72 &  0.96\\
V109 & 0.09 &  0.06 & 0.10 & 3.43  &  5.83  &  3.59 &  1.62\\
V110 & 0.51 &  0.23 & 0.10 & 4.10  &  2.09   & 6.13 &  3.44\\
V111 & 0.42 & 0.33 & 0.09 & 4.22  &  1.00  &  5.61 &  1.78\\
V112 & 0.56 & 0.38  & 0.17 &  4.00 &   1.73  &  6.03 &  4.29\\
V113 & 0.60 & 0.48 & 0.20 & 3.88  &  1.33   & 5.23  & 4.35\\
V114 & 0.66 & 0.31 & 0.19 & 3.68  &  1.61  &  5.00 &  2.34\\
V115 & 0.41 &  0.10 & 0.06 & 3.68   & 2.48  &  5.17 & 2.28\\
V116 & 0.57  & 0.46  & 0.34  & 3.84  &   1.52  &   5.43 & 4.32\\
V117 & 0.54  & 0.19  & 0.18 & 4.09  &  1.95 &   5.79  & 1.93\\
V119 & 0.59  & 0.36  & 0.23 &   3.95  &  1.73  &  5.46 & 3.81\\
V120 & 0.12  & 0.03  & 0.02  & 3.39    & 0.55    & 4.21   & 1.03\\  
V121 & 0.44  & 0.25  & 0.08  & 3.84    & 1.69    & 1.78 & 2.70\\
V122 & 0.50  & 0.29  & 0.15  & 3.94    & 1.92    & 5.83   & 3.74\\
V123 & 0.18  & 0.08  & 0.08  & 3.62    & 5.45    & 2.83   & 1.54\\
V124 & 0.09  & 0.05  & 0.07  & 3.63    & 3.83    & 2.64  & 1.38\\
V125  & 0.18  & 0.07  & 0.07  & 3.11    & 5.61    & 2.90   & 1.07\\
V140 & 0.28  & 0.16  & 0.09  & 3.62    & 1.00  &   5.38 & 1.28\\
V141 & 0.15:  & 0.23:  & 0.18:  & 5.23:    & 0.44:    & 4.54:   & 1.35:\\  
V142 & 0.35  & 0.15  & 0.11  & 5.23    & 2.94    & 5.24 & 1.98\\   
V143 & 0.70:  & 0.01:  & 0.10:  & 3.78:    & 4.00:    & 5.78:  & 2.38\\
V146(NV1) & 0.16  & 0.16  & 0.12  & 1.66  &  6.17    & 0.37 & 0.90\\
V147(NV2) & 0.13  & 0.08  & 0.07  & 3.82   & 6.00    & 2.74 & 0.92\\
V148(NV3) & 0.11  & 0.05  & 0.04  & 3.73    & 0.29    & 0.49   & 1.04\\
V149(NV4) & 0.27  & 0.12  & 0.03  & 3.31    & 0.47    & 4.32   & 1.08\\
V150(NV5) & 0.19  & 0.03  & 0.12  & 4.71    & 5.85    & 5.01 & 1.36\\
\enddata 
\end{deluxetable} 

%\clearpage

Eighteen new or suspected variables were detected in NGC~6388.
The stars labeled NV1 through NV12 seem securely established as variables.  
They would therefore become NGC~6388 variables V58 through V69,
respectively.  The nature of the suspected variables, SV1 through SV6, is
less certain.  Note that SV1 through SV4 are not the same as the suspected
variables S1 - S4 listed in \citet{Si94}.  Four of these stars are listed as being possible
type II Cepheids, based primarily upon the approximate periods indicated
by the data.  Additional data are needed to confirm this classification.
The Delta RA and Delta dec columns of Table 4 give the positions of the new
variables in arc seconds relative to the cluster center, and are on the 
system of \citet{P02}.

The Fourier and skewness parameters for the newly discovered RR Lyrae in
NGC~6388 are listed in Table~5.  Plots of $A_{21}$ versus $\phi_{21}$
and skewness versus $\phi_{21}$ are shown in Figures 7 and 8. In
Figure 7 points for known RRab and RRc stars in NGC~6388 are plotted,
using the data from Table~6 of \citet{P02}.  Although
those results were based upon standard $V$ band light curves, they
show the same pattern as do the results from our differential flux $B$
light curves.
Most of the newly discovered RR Lyrae fall clearly into the RRab or RRc
class as indicated in Table~4, but a few cases are still ambiguous.

These results clearly illustrate the power of the image subtraction method to 
detect variable stars in crowded fields. In these particular cases,
the
ability of image subtraction analysis applied to ground-based images to identify
variable stars is 
comparable to that based on much better resolved HST images, at least as regards
variables of relatively large amplitude. Only in the most crowded
areas of the cluster, within 10 arcsec of the cluster center, does the
ISIS analysis seem to do significantly worse than \citet{P03} in
detecting variable stars.  In any case, it must be noted that the seeing in our 
ground-based images was quite mediocre, ranging from $1.1\arcsec$ to $2.5\arcsec$ 
(with a typical seeing of $1.4\arcsec$). It should be remembered, though,
that the ISIS program provides light curves in the form of differential
fluxes only, and it is not possible for differential-image techniques alone to recover all 
the information that can be obtained from direct photometric techniques.  In particular, 
no mathematically rigorous algorithm exists for transforming the ISIS differential fluxes 
to magnitudes on a fundamental scale
without additional photometric information obtained by some other method.

\begin{deluxetable}{llrrl}
\tablecaption{NGC~6388 Possible New Variables \label{tbl-4}}
\tablewidth{0pt}
\tablehead{
\colhead{Variable} & \colhead{ISIS period} & \colhead{Delta RA} & \colhead{Delta 
dec} & \colhead{Type}}
\startdata
NV1(V58)    	&0.683   &-27.40 &-7.16 &RRab\\
NV2(V59)     &0.589   &6.75 &12.28 &RRab\\
NV3(V60)   &0.372   &0.00 &-16.29 &RRc\\
NV4(V61)     &0.657   &-12.70 &-5.18 &RRab\\
NV5(V62)     &0.708   &-8.73 &-0.42 &RRab\\
NV6(V63)	&2.045	 &8.52 &-2.21 &Ceph\\
NV7(V64)	&0.595   &-1.79 &-9.21 &RRab\\
NV8(V65)   	&0.395   &-4.77 &16.68 &RRc\\
NV9(V66)    	&0.350   &-10.08 &-9.16 &RRc\\
NV10(V67)   	&2.27   &-131.87 &-74.75 &Ceph\\
NV11(V68)    &0.946   &11.51 &27.76 &RRab?\\
NV12(V69)    &3.60   &3.29 &-10.84 &Ceph?\\
SV1	&~8   &4.26 &7.33 &Ceph?\\
SV2	&0.847	 &6.18 &2.90 &RRab?\\
SV3	&0.333	 &4.81 &-24.92 &RRc?\\
SV4	&~12	 &4.25 &-8.69 &Ceph?\\
SV5	&~4.5	 &0.94 &-7.18 &Ceph?\\
SV6	&~7	 &-7.78 &-24.34 &Ceph?\\
\enddata
\end{deluxetable}

%\clearpage

\begin{deluxetable}{llllllll}                                              
\tablecaption{Fourier and Skewness parameters for NGC~6388 Variables\label{tbl-5}}
\tablewidth{0pt}
\tablehead{
\colhead{Variable}  & \colhead{$A_{21}$}  & \colhead{$A_{31}$} & 
\colhead{$A_{41}$} & \colhead{$\phi_{21}$} & \colhead{$\phi_{31}$} & 
\colhead{$\phi_{41}$} & \colhead{Sk} }
\startdata
NV1  & 0.55  & 0.31  & 0.16  & 3.96    & 1.71    & 5.69 & 3.69\\  
NV2   & 0.51  & 0.35  & 0.19  & 3.77    & 1.47    & 5.47  & 3.78\\
NV3  & 0.19  & 0.13  & 0.05  & 2.48    & 5.45    & 4.23 & 1.08\\
NV4  & 0.57  & 0.38  & 0.22  & 3.84    & 1.64    & 5.73 & 4.05\\
NV5  & 0.51  & 0.26  & 0.13  & 3.85    & 1.83    & 5.46 & 3.17\\
NV7 & 0.40  & 0.30  & 0.23  & 3.99    & 1.30    & 5.17  & 1.49\\
NV8  & 0.10  & 0.07  & 0.05  & 2.95    & 5.42    & 2.64 & 0.92\\
NV9  & 0.13  & 0.08  & 0.05  & 3.16    & 5.50    & 3.07 & 1.15\\                                 
SV2   & 0.46  & 0.080  & 0.13  & 4.36    & 0.03    & 4.96 & 1.54\\
SV3  & 0.31  & 0.15  & 0.20  & 5.00    & 4.67    & 4.02 & 1.60\\
\enddata 
\end{deluxetable} 
 
%\clearpage

\section{Notes on Individual Stars in NGC~6441}
V107, V113, V115, V121--- The ISIS period that best phases our data is slightly different from the 
HST period. This is shown in Table 1.

V140 --- The \citet{P03} period does not phase our data. Our smoothest light curve is 
obtained with a period of 0.616 d. However, the light curve, as seen in Figure 
1, is very symmetrical, and, except for the relatively sharp peak, resembles 
that of an RRc variable.
A reanalysis of the WFPC2 photometry from
\citet{P03} shows that a period of 0.6141~d fits the data, although the
light curve shows the star to spend less time at minimum light than is often
the case.  Thus, it seems likely that the longer period is correct. Note that at 
present the longest confirmed period for an RRc variable in a globular cluster 
is about 0.56~d, with only one candidate RRc having recently been suggested 
with a period longer than 0.6~d (Contreras et al. 2005). Because
the Fourier and skewness parameters for V140 do not give a clear
classification, further data are
needed to confirm whether V140 is an RRab or RRc star.  

V141 --- Our smoothest light curve is 
obtained with a period of 0.457~d, as shown in Figure 1.
However, the resulting light curve is unusual 
looking.
A period of 0.847~d gives almost as good a light curve and is more
consistent with the period in \citet{P03}.  Thus, the longer
period has been adopted.

V142 --- The HST period phases our data well. However, the resulting light curve is 
unusual looking.

NV1 (V146) --- We could not find a period that phases our data well. The best 
period we could determine is 0.402 day, which may suggest an RRc variable.
We
note that, among Oosterhoff type II clusters such as M15, a period near
0.402 day and a light curve showing scatter are sometimes indicators
that a star is a double-mode pulsator.  However, no double-mode RR Lyrae
stars have yet been discovered in NGC~6441 and, since NGC~6441 seems to 
contain some RRc stars with periods longer than 0.402 day, it is not
clear whether one might expect that double-mode RR Lyrae stars in
that cluster would have first overtone mode periods near 0.402 day.

NV4 (V149) and NV5 (V150)--- These variables have periods of 0.557~d and 0.529~d, again 
relatively long for RRc-type variables.  However, the differential flux light curves
are more symmetric than is typical of RRab variables, which would suggest
an RRc classification.  Long-period RRc variables, while exceedingly
rare in globular clusters in general (Catelan 2004 and references therein),
have previously been found in both NGC~6388 and NGC~6441 \citep{P01,P02,P03}.

\begin{figure}[t]
  \figurenum{5}
  \epsscale{0.85}
\plotone{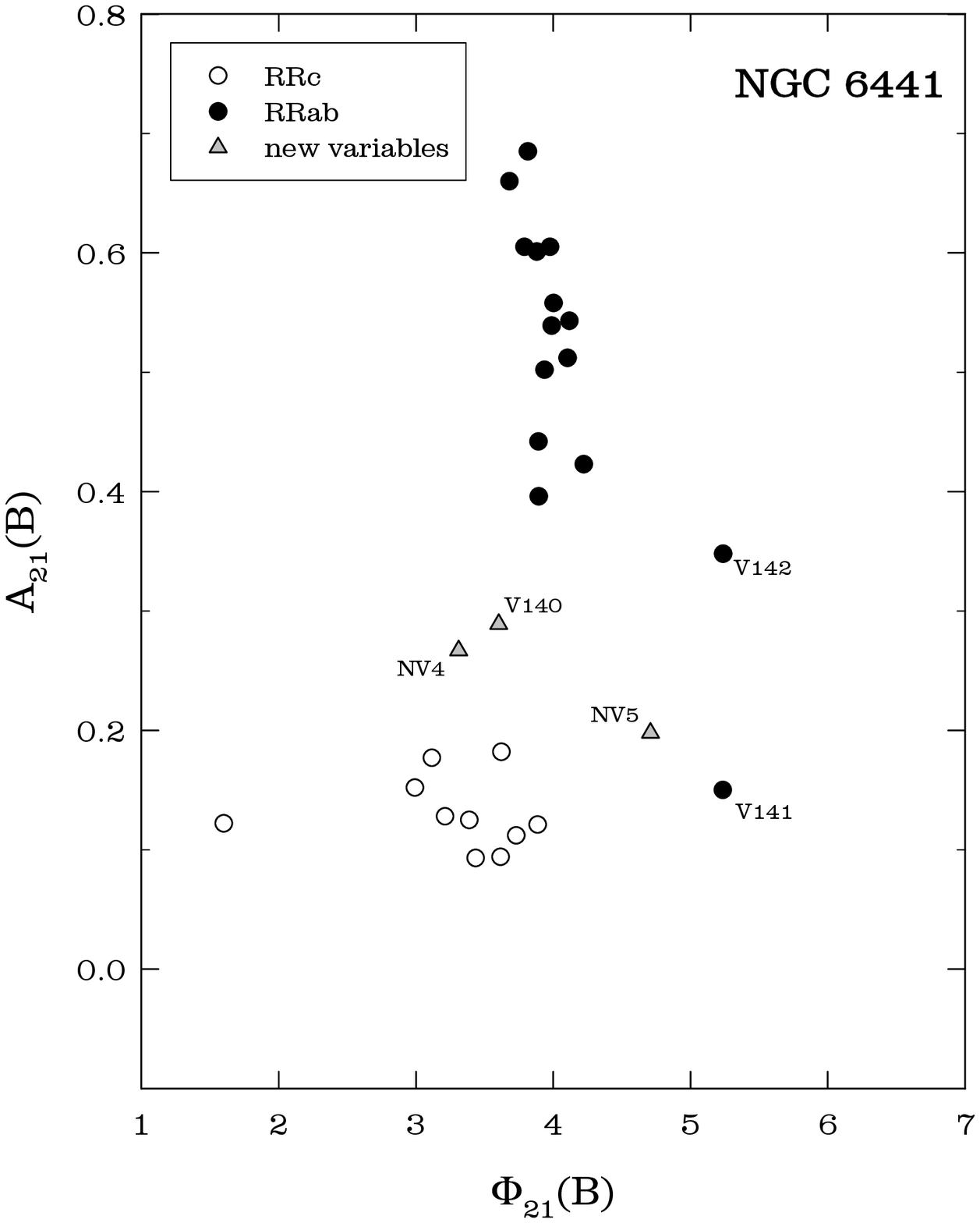}
  \caption{Plot of the Fourier parameter $A_{21}$ versus $\phi_{21}$
  for RR Lyrae stars in NGC 6441.   
      }
      \label{fig5}
\end{figure}

\section{Discussion}

The ISIS analysis of the ground-based observations of NGC 6441, in spite
of seeing ranging from $1.1\arcsec$ to $2.5\arcsec$, rediscovered virtually all
of the RR Lyrae stars and Cepheids catalogued by \citet{P03}.  ISIS also
identified a few possible variable candidates not appearing in the 
Pritzl et al. catalog.
This confirms the utility of ISIS for identifying and classifying the brighter
variable stars in crowded fields. On the other hand, 
ISIS does not by itself provide light curves on
a standard photometric system, as do reduction routines such as {\sc daophot}, 
nor does it give astrometric positions as accurate as those obtained from
WFPC2.

Including the five newly discovered variables, we
find that NGC~6441 is now known to contain 68 probable RR Lyrae variables.
Accepting NV1, 2, and 3 as RRc variables, and taking account of the
uncertainty Bailey type of V140, NGC~6441 contains 42 known
RRab stars and 23 known RRc stars, changing the ratio of RRc to all RR
Lyrae stars from the value of 0.33 found in \citet{P03} to 0.35.  The resultant
mean periods are 0.759~d and 0.392~d for the RRab and RRc stars, respectively.
If NV4 and NV5 are also actually RRc variables, then the total of RRc
stars increases to 25 and their mean period increases to 0.404~d.  In
either case, NGC~6441 remains among the clusters with the largest values
of $\langle P_{ab}\rangle$ and $\langle P_{c}\rangle$.

For NGC~6388, five of the new variables seem to be clearly RRab stars
and three are RRc stars.  Adding these variables to those in \citet{P02}
gives a total of 22 probable RR Lyrae stars.  Of these, nine are
RRab stars and 11 are RRc stars (if, as \citet{P02} suggest,
V26 and V34 are excluded as nonmembers). These totals do not include the variables
listed as questionable c or ab type stars in Table 3 of
\citet{P02}. It is probable that many of the questionable stars are in fact
RR Lyrae variables, as indicated in \citet{P02}, but for various
reasons the light curves of these stars were noisy or incomplete.
 If all of the additional stars 
listed as ``c?" or ``ab?" variables
in Table 4 are included, the number of RRc stars goes up by 7 and the
number of RRab variables goes up by 2. From the
confirmed RRab and RRc variables, \citet{P02} obtained a ratio of RRc to
total RR Lyrae stars of 0.67 (or 0.71 if v26 and V34 are included).  
The new discoveries revise this ratio to 0.55 (without V26 and
V34) or 0.59 (with V26 and V34).  The resultant mean periods are
$\langle P_{ab}\rangle = 0.676$~d (excluding all questionable RRab stars) and 
$\langle P_{c}\rangle  = 0.364$~d (with V26 and V34) or 0.387~d
(without V26 and V34).  Because the new variables make a
significant addition to the RR Lyrae inventory for NGC~6388,
we show a revised histogram over period in Figure 9. V26 and V34 have
been excluded from this histogram as possible field stars, but all other probable
RR Lyrae stars in Table 3 of \citet{P02} have been included.
The identification of additional probable and 
suspected type II Cepheids
in NGC~6388 confirms its status as a cluster rich in such variables.
This further strengthens the unusual status of NGC~6388 and NGC~6441
as metal-rich globular clusters also rich in type II Cepheids.

\begin{figure}[t]
  \figurenum{6}
  \epsscale{0.85}
\plotone{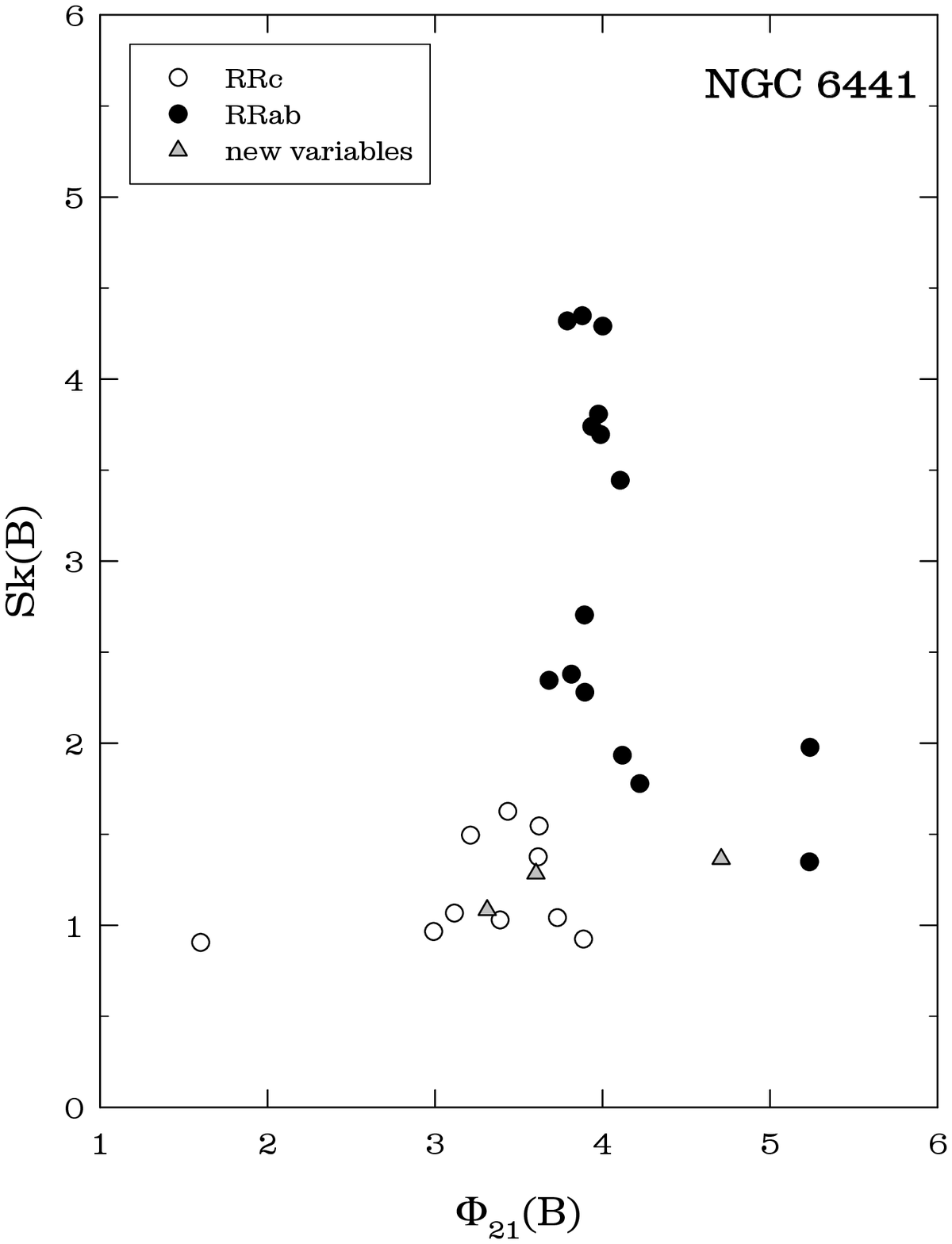}
  \caption{Plot of the skewness parameter Sk versus the Fourier phase
  parameter $\phi_{21}$ for variable stars in NGC 6441. The RR Lyrae stars
  of uncertain classification, NV4, NV5, and V140, are indicated by
  triangles. 
      }
      \label{fig6}
\end{figure}

The periods of the RRc stars in NGC~6388 and NGC~6441
can be transformed to their fundamental mode equivalents by adding 0.128 to 
the logarithms of their periods.  The resultant histograms for RR Lyrae
stars in these two clusters are shown in Figure 10.  In the case of
NGC~6441, we plot V118 as an RRab star, although \citet{P03} noted
that it might possibly be a type II Cepheid.  The newly found variables
V146, V147, and V148 have been included.  The fundamentalized 
histogram for NGC~6388
climbs toward the short period end, indicating that the hotter side of
the RR Lyrae instability strip is more populated than the cooler side.
Although the histogram for NGC~6441 shows some peaks, no such overall
trend is clearly discernable. Color-magnitude diagrams of the horizontal
branches of NGC~6388 and NGC~6441 \citep{P03, B04} show that the density
of stars on the blue extensions to the horizontal branch in both cases
declines toward the cool side of the instability strip, eventually
increasing again when a strong concentration of red horizontal branch
stars is reached.  If this decline takes place within the instability
strip in the case of NGC~6388 but mostly to the cool side of the
instability strip in the case of NGC~6441, then the existence of a trend
in the NGC~6388 histogram but not in that of NGC~6441 can be understood.
For further discussion on the relation between the star distribution
in the HR diagram and the resulting period distribution, the reader is
referred to \citet{R89}, \citet{Cat04}, and \citet{Cas05}.

%                                                One column figure
%----------------------------------------------------------- Fig. 7  
\begin{figure}[t]
  \figurenum{7}
  \epsscale{0.85}
\plotone{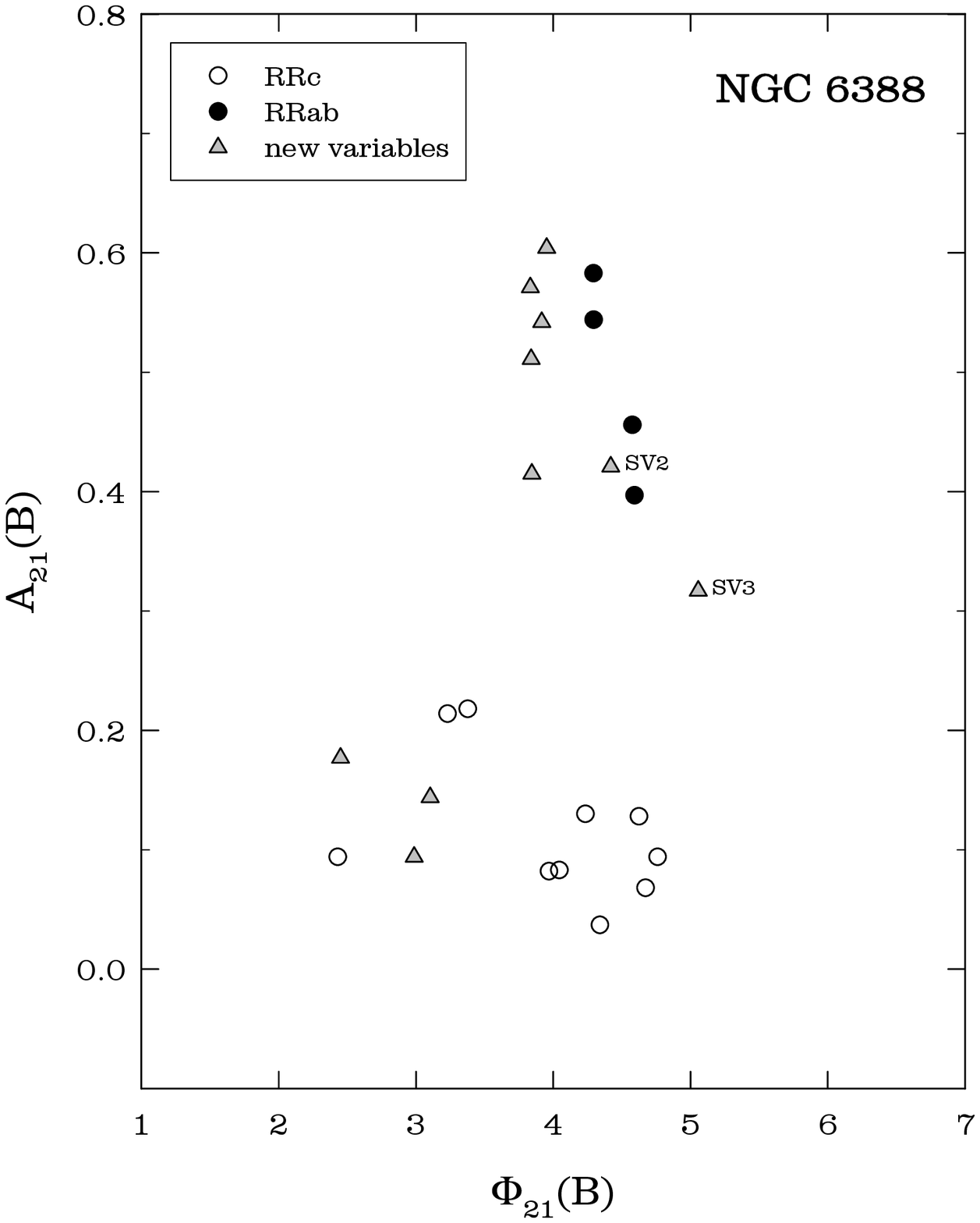}
  \caption{The Fourier $A_{21}$ parameter is plotted against $\phi_{21}$
  for the newly discovered RR Lyrae stars in NGC 6388. The filled and
  open circles indicate RRab and RRc stars in NGC~6388, respectively,
  plotted with data from Table~6 of \citet{P02}.
      }
      \label{fig7}
\end{figure}

Although their metallicities are much higher than those of
canonical Oosterhoff type II clusters such as M15 (NGC 7078) or M68
(NGC 4590),
the mean periods of RR Lyrae stars in NGC~6388 and NGC~6441 are as large
as, or larger than, those in Oosterhoff type II systems.  We can compare
the fundamentalized histograms of Figure 10 with those of M15 and
M68 -- see Figure 1 in \citet{Cas04}.  NGC~6441 and, to a lesser
extent, NGC~6388 have a higher proportion of RR Lyrae stars with
periods longer than 0.8 days.  Otherwise, the differences are not
greater than are seen among the histograms of more ordinary clusters.
The peak toward the shorter period end of the NGC~6388 histogram is
similar to that seen in the histogram of M68.  M15 also shows a peak
toward shorter periods, but, like NGC~6441, the Oosterhoff type II
cluster M2 (NGC 7089) shows a relatively flat distribution of RR Lyrae periods.
In this sense, it is clear that neither NGC~6388 nor NGC~6441
show sharply peaked period distributions to the same extent that is seen
in the case of the Oosterhoff type I cluster M3 (NGC 5272) \citep{R89,Cat04}.
\citet{P02} did note that there was one Oosterhoff type II globular
cluster that shared some of the peculiarities of NGC~6388 and NGC~6441.
The unusual globular cluster $\omega$ Centauri, like NGC~6388
and NGC~6441, contains some RRab and RRc stars of especially long period.
In the case of $\omega$ Cen the long period RR Lyrae stars are, however,
accompanied by a shorter period RR Lyrae population, so that the mean
period of RRab stars in $\omega$ Cen is shorter than in 
NGC~6388 or NGC~6441.

\begin{acknowledgements}
H.A.S. thanks the Center for the Study of Cosmic Evolution and 
the National Science 
Foundation for partial support of this work under
grant AST-0205813. Support for M.C. was provided by Proyecto FONDECYT Regular 
No. 1030954.
Support for B.J.P. was provided through a National Science
Foundation CAREER award, AST-9984073. The observations for the NGC~6441
variables were obtained in part with the NASA/ESA Hubble Space Telescope
under program SNAP8251.  The Space Telescope Science Institute is 
operated by the Association of Universities for Research in Astronomy,
Inc. under NASA contract NAS 5-26555.
\end{acknowledgements}

%                                                One column figure
%----------------------------------------------------------- Fig. 8  
\begin{figure}[t!]
  \figurenum{8}
  \epsscale{0.85}
\plotone{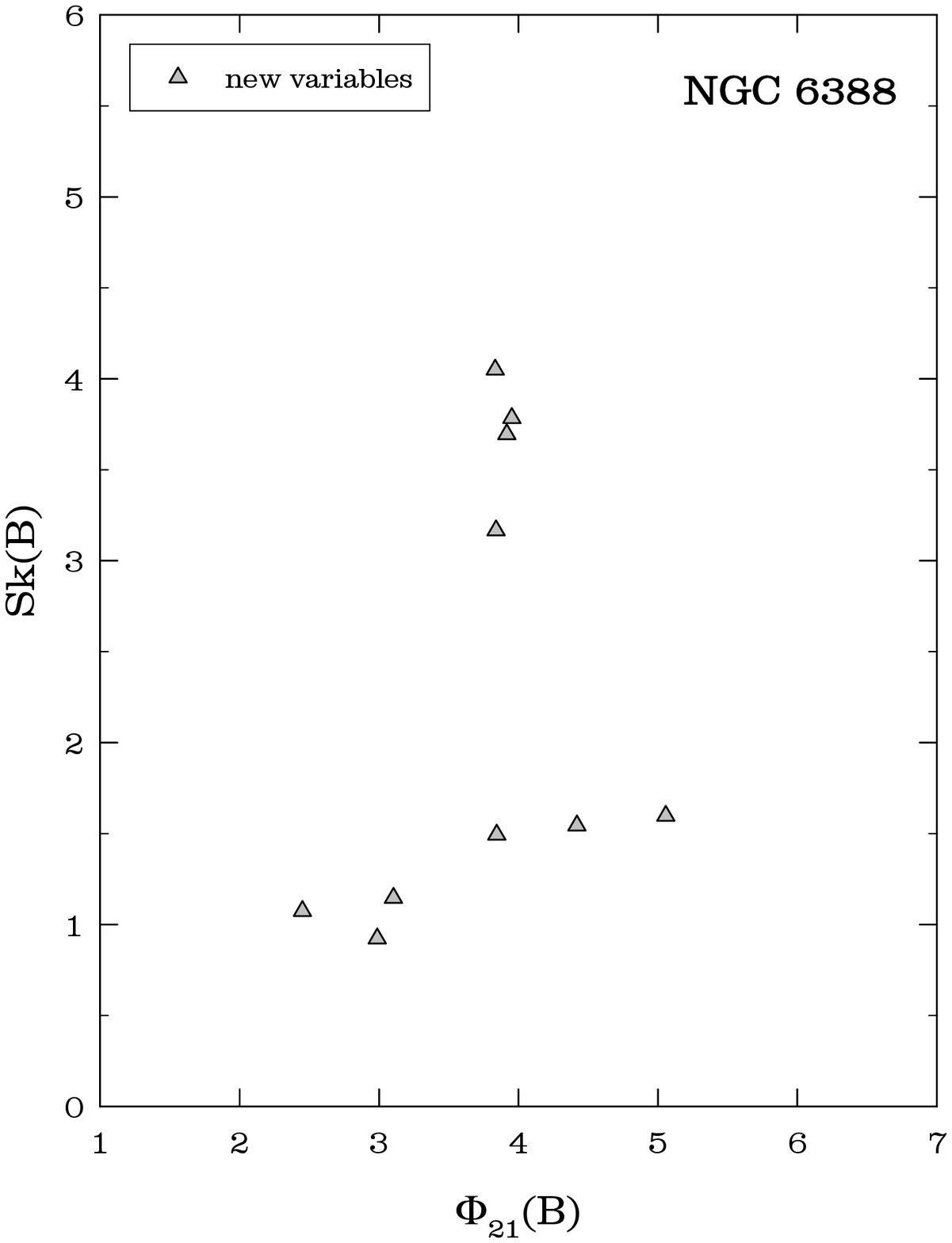}
  \caption{The skewness parameter Sk is plotted against $\phi_{21}$
  for the newly discovered RR Lyrae stars in NGC~6388.
      }
      \label{fig8}
\end{figure}

%                                                One column figure
%----------------------------------------------------------- Fig. 9  
\begin{figure}[t]
  \figurenum{9}
%  \epsscale{0.85}
\plotone{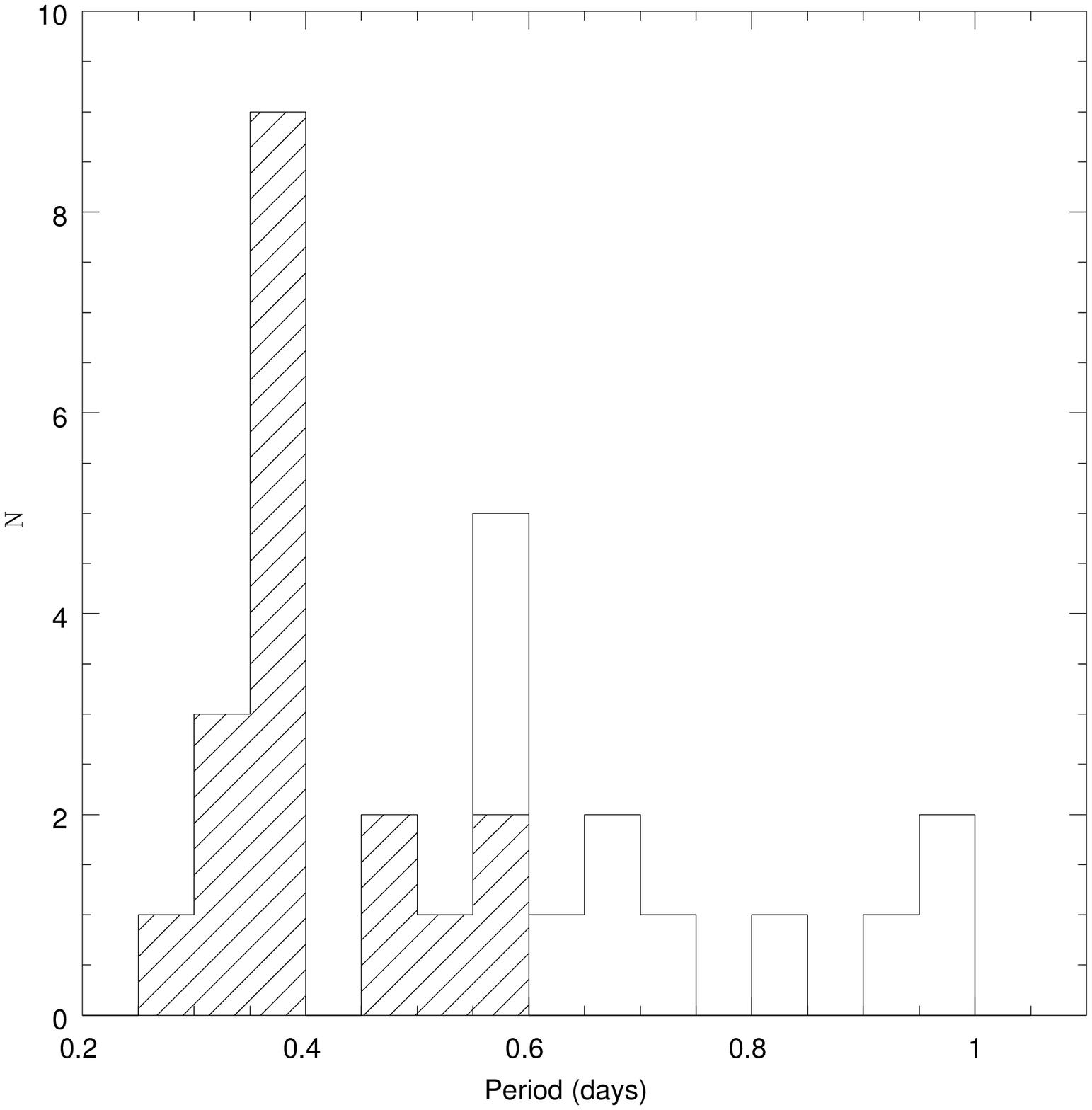}
  \caption{Histogram over period for RR Lyrae stars in NGC~6388. The hatching
  indicates the histogram for RRc stars.
      }
      \label{fig9}
\end{figure}

\begin{figure}[t]
  \figurenum{10}
%  \epsscale{0.85}
\plotone{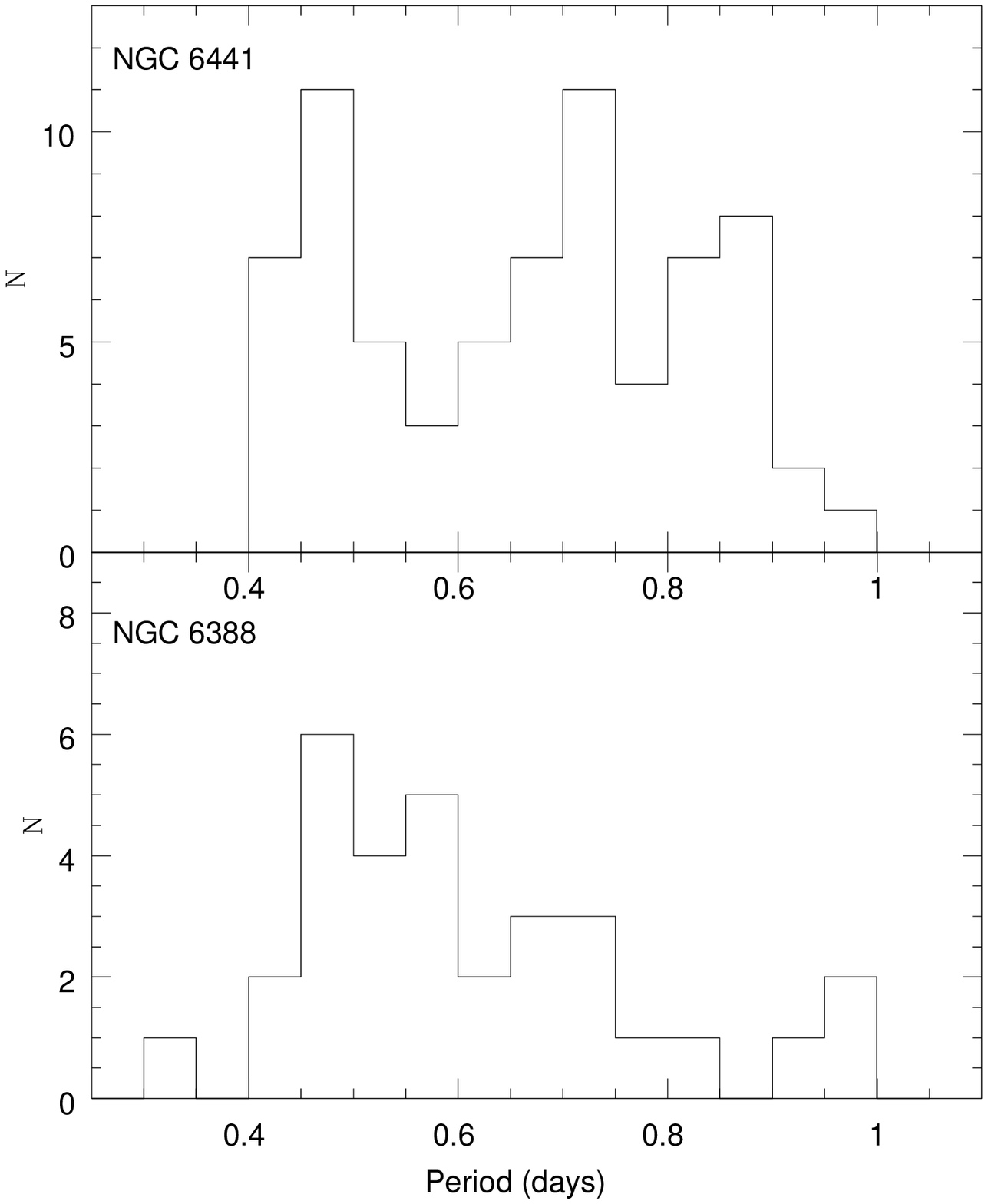}
  \caption{Histogram over period for RR Lyrae stars in NGC~6388 and NGC~6441, 
  after the periods of the RRc variables have been converted to their
  fundamental mode equivalents.
      }
      \label{fig10}
\end{figure}

\end{document}